\documentclass{article}

\PassOptionsToPackage{numbers, compress}{natbib}

\usepackage[preprint]{nips_2018}

\usepackage[utf8]{inputenc} 
\usepackage[T1]{fontenc}    
\usepackage{hyperref}       
\usepackage{url}            
\usepackage{booktabs}       
\usepackage{amsfonts}       
\usepackage{graphicx}
\usepackage{xcolor}
\usepackage{float}
\usepackage{amsmath}
\usepackage{siunitx}
\usepackage{multirow}

\DeclareMathOperator*{\argmin}{arg\,min}

\title{
    Low-latency job scheduling with preemption \\
    for the development of deep learning
}

\author{
    Hidehito Yabuuchi\thanks{
        Work done during an internship at Preferred Networks, Inc.
    } \\
    The University of Tokyo \\
    \texttt{yabuuchi@os.ecc.u-tokyo.ac.jp} \\
    \And
    Daisuke Taniwaki \hspace{1.2em} Shingo Omura \\
    Preferred Networks, Inc. \\
    \texttt{\{dtaniwaki, omura\}@preferred.jp} \\
}

\begin{document}

\maketitle

\begin{abstract}
    One significant challenge in the job scheduling of computing clusters for the development of deep learning algorithms is
    the efficient scheduling of \textit{trial-and-error (TE)} job, the type of job in which the users seek to conduct small-scale experiments
    while monitoring their processes.
    Unfortunately, the existing job schedulers to date do not feature well-balanced scheduling for the mixture of TE jobs and \textit{best-effort (BE)} jobs,
    or they can handle the mixture in limited situations at most.
    To fill in this niche, we propose an algorithm that can significantly reduce the latency of TE jobs in versatile situations
    without greatly elongating the slowdown of the BE jobs.
    Our algorithm efficiently schedules both TE and BE jobs by selectively preempting the BE jobs that can be, when the time comes,
    resumed without much delay.
    In our simulation study with synthetic and real workloads,
    we were able to reduce the 95th percentile of the \textit{slowdown rates} for the TE jobs in the standard FIFO strategy by 96.6\%,
    while compromising the median of the BE slowdown rates by only 18.0\% and the 95th percentile by only 23.9\%.
\end{abstract}

\section{Introduction}
\label{sec:intro}

Efficient resource management of computing clusters is in high demand in research institutes and companies across the world these days,
especially due to the recent explosive development of deep learning (DL) algorithms
in various fields such as image recognition and natural language processing.
In general, the development of DL requires large computing resources including GPUs.
The resource management on a cluster is often conducted by a job scheduler.
The job scheduler is responsible for selecting the nodes with available resources and allocating the resources to the submitted jobs
that are deemed to have high priority.
The value of the scheduler is generally determined by its ability to process large throughput and its ability to minimize the latency,
which is the sheer length of the time that the user must wait until the cluster begins processing the submitted job.

In the development of DL, the type of jobs that play a pivotal role is \textit{trial-and-error (TE)} jobs,
in which the users conduct small-scale experiments on a trial basis.
Indeed, many jobs submitted for the development of DL are dedicated for the debugging and the testing of the prototype algorithms,
and it is typical to conduct such small experiments prior to official submission and production.
The users often want to frequently monitor the learning curves of the prototypes in order to save time for exploring numerous other options.
Meanwhile, other jobs can be executed in the \textit{best-effort (BE)} manner.
In fact, our analysis on the private cluster at the authors' institution suggests
that the TE jobs have accounted for approximately 30\% of all jobs submitted to the cluster in six months.

Unfortunately, most scheduling algorithms to date can handle the mixture of TE and BE jobs in certain situations at most,
depending on the choice of the software architectures.
Still other scheduling algorithms go only so far to allocate a separate fixed resource for the TE jobs.
Big-C \cite{Chen17} is a scheduler that falls into the first category, and it reduces the latency of the short jobs
via fine-grained resource management of lightweight virtualization.
Big-C algorithm works by preempting the job with the longest remaining execution time and allocating the resulting surplus to the short jobs.
However, to the best of our knowledge, there are no production-level methodologies to date that can multiplex the GPUs with virtualization.
Big-C is also not applicable to distributed DL algorithms, which communicate between containers.
Optimus \cite{Peng18} is a job scheduler specialized for the handling of the training jobs of deep neural networks (DNNs) on the parameter-server architectures
(e.g., TensorFlow \cite{Abadi16} and MXNet \cite{Chen15}), and it is compatible with dynamic resource allocation.
Optimus ranks the priority of the submitted jobs by probing their learning speeds using a variety of resource allocation patterns.
This dynamic feature of Optimus, at the same time, makes Optimus incompatible with ChainerMN \cite{Akiba17} and PyTorch \cite{Paszke17},
which both adopt all-reduce architecture for scalability.
Gandiva \cite{Xiao2018} is another job scheduling framework that improves the latency and efficiency of the training jobs of DNNs.
Exploiting the iterative nature of the training jobs, Gandiva efficiently time-slices the GPUs across multiple jobs,
and introspects the performance of each job and determines whether it should be migrated to different GPU.
However, Gandiva's introspection requires a very specific configuration of DL frameworks, and its use is limited.
Reservation-based schedulers such as Hawk \cite{Delgado15}, on the other hand,
take the strategy of reserving a separate portion of a cluster in order to guarantee the immediate scheduling of short jobs.
Given highly diverse workloads, however, it is often a challenging task to find the optimal reservation factor.

In this paper, we take the strategy of systematically suspending a selected set of BE jobs in favor of the user-specified TE jobs.
Our proposed algorithm can handle any DL jobs that can be suspended, and it can be used in versatile situations.
We also take special care to make sure that the BE jobs are not \textit{neglected}.
By selectively preempting the BE jobs for which the scheduler can re-schedule its execution in relatively short time,
our algorithm makes sure not to greatly increase the overall slowdown of the BE jobs.
As we will show in Section \ref{sec:preempt-alg}, this can be done at relative ease by observing the amount of the resource demanded by the BE jobs.
In fact, unlike Big-C and Optimus, our algorithm does not have to estimate the execution time of the jobs.
This is an important feature of our algorithm because the execution time are generally difficult to estimate \cite{Delgado18},
which is especially the case for the DL jobs whose execution time are sensitive to the choice of the hyper-parameters.
We also guarantee that each BE job is not starved by limiting the number of preemption to a fixed number of times.

\section{System model}
\label{sec:sysmodel}

In this section, we describe the system model assumed in our algorithm, along with the required nomenclatures and definitions.
For our algorithm, the users must first specify the type of the jobs to be submitted, either TE or BE,
along with the type and amount of the resource the users want to request for each job.
The scheduler is given the permission to suspend BE jobs when needed.
Suspended BE jobs are placed back on the top of the job queue, and the surplus resource that becomes available upon the suspension is allocated to the TE jobs.

Our algorithm prompts the user for still another piece of information.
Some jobs require the time for suspension processing (e.g., writing data back to persistent storage) before being suspended.
Large DL jobs that process large model on RAM tend to require a long time for the suspension processing.
We therefore allow a \textit{grace period (GP)} of user-specified length for each suspension prompt.
In order to make room for the GP, we design the system so that it sends a signal to a job well before its preemption.
If a user is willing to permit \textit{rewinding} of a job (e.g., for one that periodically takes snapshots), the user may set the length of GP to zero.

For the ease of description, in this paper, we assume that we can use three types of resources: CPU, RAM, and GPU.
The extension of our theory to other types of resource should be straightforward.

Also, we assume that each job consists of a single task, because unlike big-data processing (e.g., MapReduce jobs),
a typical DL job tends not to have multiple tasks that form a DAG.

\section{Proposed preemption algorithm}
\label{sec:preempt-alg}

We built our preemption algorithm on the FIFO principle,
which is widely used in production (e.g., YARN \cite{Vavilapalli13}, Spark standalone mode \cite{Zaharia10}, and Kubernetes \cite{Burns16}),
so that we can easily integrate our algorithm into the existing frameworks.

When a TE job arrives at a job queue, one or more BE jobs are suspended to make room for the incoming TE job if the resource is insufficient for the TE job.
To observe the FIFO principle, the preempted jobs are to be placed back on the top of the queue.

\subsection{Effect of the preemption on the BE jobs}
\label{subsec:preempt-effect}

Our algorithm carefully selects the BE jobs to be preempted so that the overall completion time of BE jobs are not excessively prolonged.
Big-C \cite{Chen17} was built on the belief that the preemption has little effect on the slowdown of the long-duration jobs.
This belief, however, is not necessarily true.
Indeed, the preemption process on its own does not have much effect on the completion time of the preempted BE jobs
because they are always placed back on the top of the queue upon the suspension and they are to be re-scheduled without much delay.
In our experiment as well, the intervals between preemptions and re-scheduling were less than five minutes in most cases (Table \ref{tbl:resched-intv}).
This is not to say, however, that the other BE jobs waiting in the queue are unaffected;
We observed that their processing time is strongly affected by the preemption.
If a BE job that demands large resource is preempted, the \textit{head-of-line blocking} will occur with high probability.
In other words, the scheduler must wait for a long time to re-schedule the execution of the \textit{large} BE job,
and other BE jobs must wait for even longer time.
Our algorithm therefore preempts the BE jobs for which the scheduler can readily find the room for re-scheduling.
In this way, we can reduce the additional waiting time to be incurred for other BE jobs.

On the other hand, preempting too small a BE job prolongs the overall time required for the completion of BE jobs.
If the preemption of one BE job cannot make enough room for the incoming TE job, the scheduler will have to preempt still another BE job.
Our algorithm therefore makes an effort to also reduce the sheer number of preemptions,
and hence the total time-loss incurred by the re-scheduling intervals.

\subsection{Proposed algorithm}
\label{subsec:alg}

Motivated by above observations and dilemmas,
we designed a preemption algorithm, \textit{Fitting Grace Period Preemption (FitGpp)}, based on the following four strategies.

\begin{figure}[ht]
    \centering
    \includegraphics[width=0.45\linewidth]{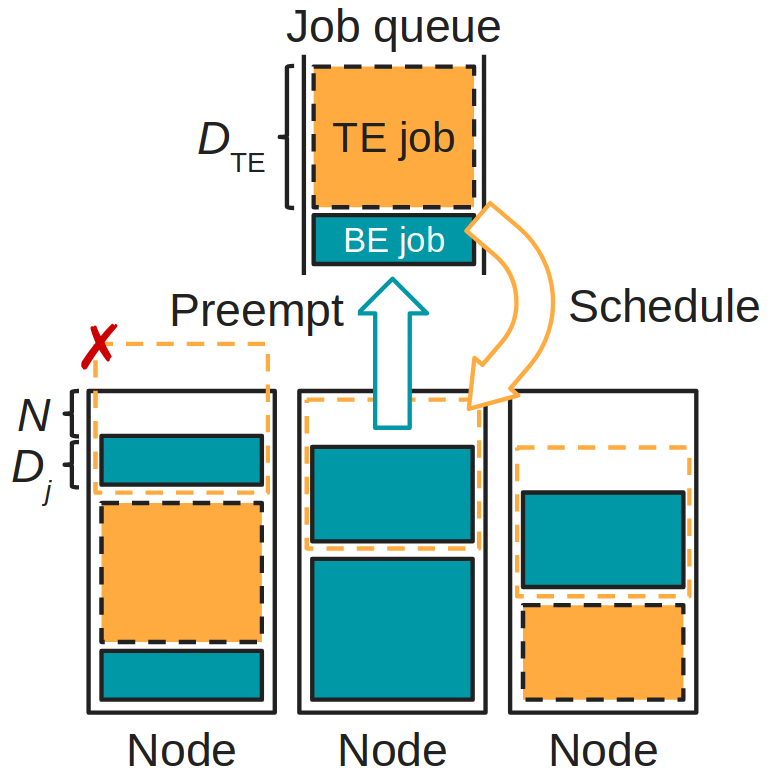}
    \caption{
        The overview of our FitGpp algorithm.
        When the unallocated resource is insufficient for the incoming TE job (the light orange cell with dashed border line),
        the scheduler preempts one or more BE jobs (the dark blue cells with solid border lines).
        FitGpp selects neither of the two BE jobs in the left node because they cannot offer enough resource for the TE job.
        FitGpp selects the BE job with the minimum score (Eq. \ref{eq:fitgpp-score})
        among the list of BE jobs that have not reached the maximum preemption limit $P$.
    }
    \label{fig:fitgpp}
\end{figure}

\begin{description}
\item[Minimizing the re-scheduling intervals.]
At all time, our FitGpp algorithm selects a BE job with small resource demand.
When $C, R, G \in \mathbf{R}^+$ respectively represent the demands made by a certain job for CPU, RAM, and GPU,
our FitGpp measures the \textit{size} of the resource demand using the following formula:

\begin{equation}
    \mathrm{Size}([C, R, G]) := \sqrt{
        \left( \frac{ C }{ C_{\mathrm{capacity}} } \right)^2 +
        \left( \frac{ R }{ R_{\mathrm{capacity}} } \right)^2 +
        \left( \frac{ G }{ G_{\mathrm{capacity}} } \right)^2
    }
\end{equation}

where $X_\mathrm{capacity}$ represents the capacity of the resource $X$ in the node.
Note that this formulation is invariant under the choice of the measurement scale (e.g., 4 CPUs vs. 64 GB RAM).

\item[Minimizing the number of preemptions.]
To reduce the number of unnecessary preemptions,
our FitGpp algorithm also prioritizes the preemption of a BE job that can offer enough resource for the incoming TE job on its own.
More precisely, FitGpp preempts the BE job only if

\begin{equation}
    D_{\mathrm{TE}} \le D_{\mathrm{BE}} + N \quad (\text{element-wise inequality})
\end{equation}

holds, where $D_X$ represents the demand vector $[C_X, R_X, G_X]$ for the $X$ job
and $N$ represents the vector $[C_N, R_N, G_N]$ describing the size of the respective unallocated resources in the node on which the BE job is running.

\item[Minimizing the preemption-incurred time loss.]
As we described in Section \ref{sec:sysmodel}, our system model grants a grace period (GP) to any BE job upon its suspension,
the period during which the job is allowed to perform a suspension processing.
It is obviously not preferable to preempt a BE job with too long a GP,
because the length of the GP affects the time until the execution of the incoming TE jobs.
Because large BE jobs tend to require long GPs, our FitGpp algorithm preferentially preempts the BE jobs with short GPs.

\item[Avoiding the starvation of BE jobs.]
In order to ensure that BE jobs are not preempted infinitely many times,
our FitGpp algorithm does not preempt any given BE job more than a fixed number of times, $P$.
\end{description}

In order to formalize the four strategies above,
our FitGpp algorithm evaluates the following score for each job $j$:

\begin{equation}
\label{eq:fitgpp-score}
    \mathrm{Score}(j) := \frac{ \mathrm{Size}(D_j) }{ \max_{ j \in \mathcal{J} }{ \mathrm{Size}(D_j) } }
    + s \times \frac{ \mathrm{GP}_j }{ \max_{ j \in \mathcal{J} }{ \mathrm{GP}_j } }
\end{equation}

where $D_j$ is the resource demanded by $j$, and $\mathcal{J}$ represents the set of all currently running BE jobs.
The parameter $s$ determines the importance of the GP relative to the resource demand.
We conducted the sensitivity analysis with respect to $s$ in Section \ref{subsec:sens-anal}.
FitGpp then chooses the BE job to be preempted (denoted by $j^*$) using the following rule (Fig. \ref{fig:fitgpp}):

\begin{equation}
     j^*  =  \argmin \left\{ \mathrm{Score}(j) \mid D_{\mathrm{TE}} \le D_j + N \ \land \ \mathrm{PreemptionCount}_j < P \right\}
\end{equation}

where $\mathrm{PreemptionCount}_j$ represents the number of times that the job $j$ has been preempted.
If there is no running BE job that meets the condition, FitGpp preempts a random BE job.
For large clusters with many nodes, however, this type of situation is rare.
In fact, this never happened in our experiments described in Section \ref{sec:eval}.

Note that the criteria of preemption used in our FitGpp algorithm is not dependent on the execution time,
so that it is not affected by the algorithm's ability to estimate the execution time.
This is an important feature of FitGpp because the estimation of execution time is generally hard \cite{Delgado18};
this is especially the case for the DL jobs, whose execution time are sensitive to the choice of the dataset and the hyper-parameters.

\section{Evaluation}
\label{sec:eval}

\subsection{Setup}
\label{subsec:eval-setup}

We evaluated our FitGpp algorithm using a simulator we developed for this purpose.
The simulated environment consisted of 84 nodes, each having 32 CPUs, 256 GB RAM, and 8 GPUs.
This configuration is the same as the private cluster for the development of DL at the authors' institution.
In the simulation, the job scheduler decides resource allocation at every simulated minute.

We compared FitGpp against (non-preemptive) vanilla FIFO, \textit{Longest Remaining Time Preemption (LRTP)}, and \textit{RAND}.
LRTP is the algorithm used in Big-C \cite{Chen17} and it preferentially preempts the job with the longest remaining execution time.
We simulated LRTP on the assumption that it can perfectly predict the execution time.
RAND is a strategy that preempts a randomly selected running BE job.
Both LRTP and RAND continue the preemption process until they can prepare enough resource for the incoming TE job.
We compared the performance of the algorithms based on the \textit{slowdown rate} computed by the formula

\begin{equation}
    1 + \frac{ \mathrm{WaitingTime} }{ \mathrm{ExecutionTime} }.
\end{equation}

In general, the smaller the slowdown rate, the better the performance.

\subsection{Experiment with synthetic workloads}
\label{subsec:synth-workload}

\begin{figure}[ht]
    \centering
    \includegraphics[width=\linewidth]{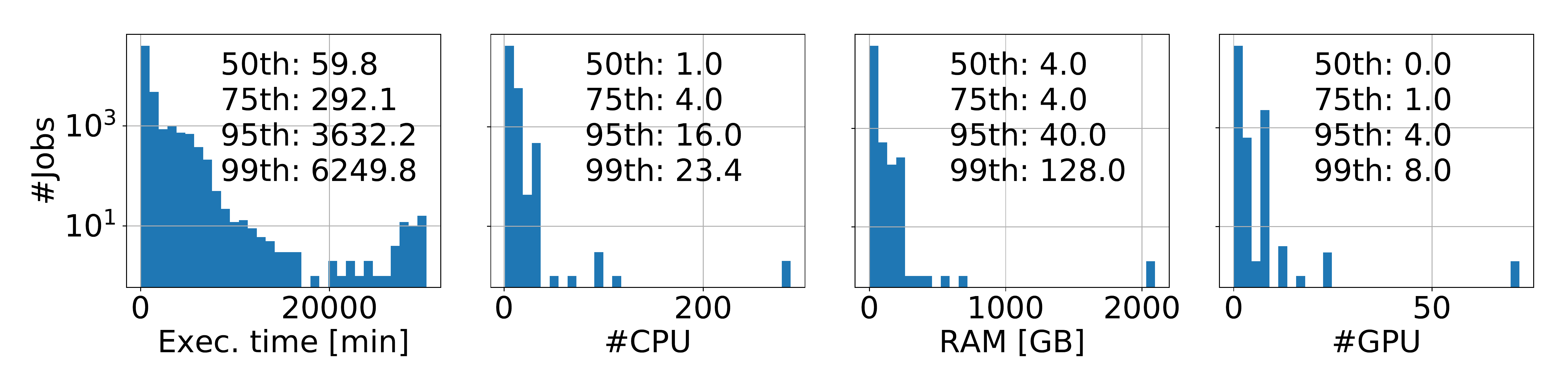}
    \caption{Statistics of jobs on the cluster at the authors' institution.}
    \label{fig:workload-stats}
\end{figure}

In order to generate realistic workloads, we analyzed a trace of the cluster at the authors' institution.
The number of jobs that lasted more than \SI{180}{s} in the trace over six months period was approximately 50,000.
Fig. \ref{fig:workload-stats} shows its brief statistics.
To create a realistic sequence of synthetic workloads, we approximated the empirical distributions of (1) execution time, (2) CPU, (3) RAM, and (4) GPU
for both TE jobs and BE jobs with separate normal distributions, and artificially generated typical jobs from their truncated versions.
The means of the fitted normal distributions for the execution times of the TE jobs and the BE jobs were respectively \SI{5}{min.} and \SI{30}{min}.
We truncated these distributions at \SI{30}{min.} and \SI{24}{hours}, in this order.

For the lengths of GPs, we prepared the normal distribution with the mean of \SI{3}{min.} and truncated the distribution at \SI{20}{min}.
We set the length of GPs at such large values for the following three reasons:
(1) typical DL jobs tend to accompany large data to store before the suspension,
(2) the data often requires preprocessing step for the storage, such as serialization, and
(3) we expect the developers of DL algorithms to specify long GPs because prematurely suspended job is destined to fail.

We evaluated FitGpp with eight sets of generated workloads, each consisting of $2^{16}$ jobs with 30\% of them being TE.
The jobs were submitted at such a rate that the cluster load (the ratio of the total resource demand relative to the capacity)
would be kept at 2.0 if they were scheduled by FIFO.
For the evaluation of RAND, we repeated the same experiment four times and reported the average statistics.
The $P$ value was set to 1 in this experiment.
In other words, each BE job is not preempted more than once.

\begin{figure}[ht]
    \centering
    \includegraphics[width=0.75\linewidth]{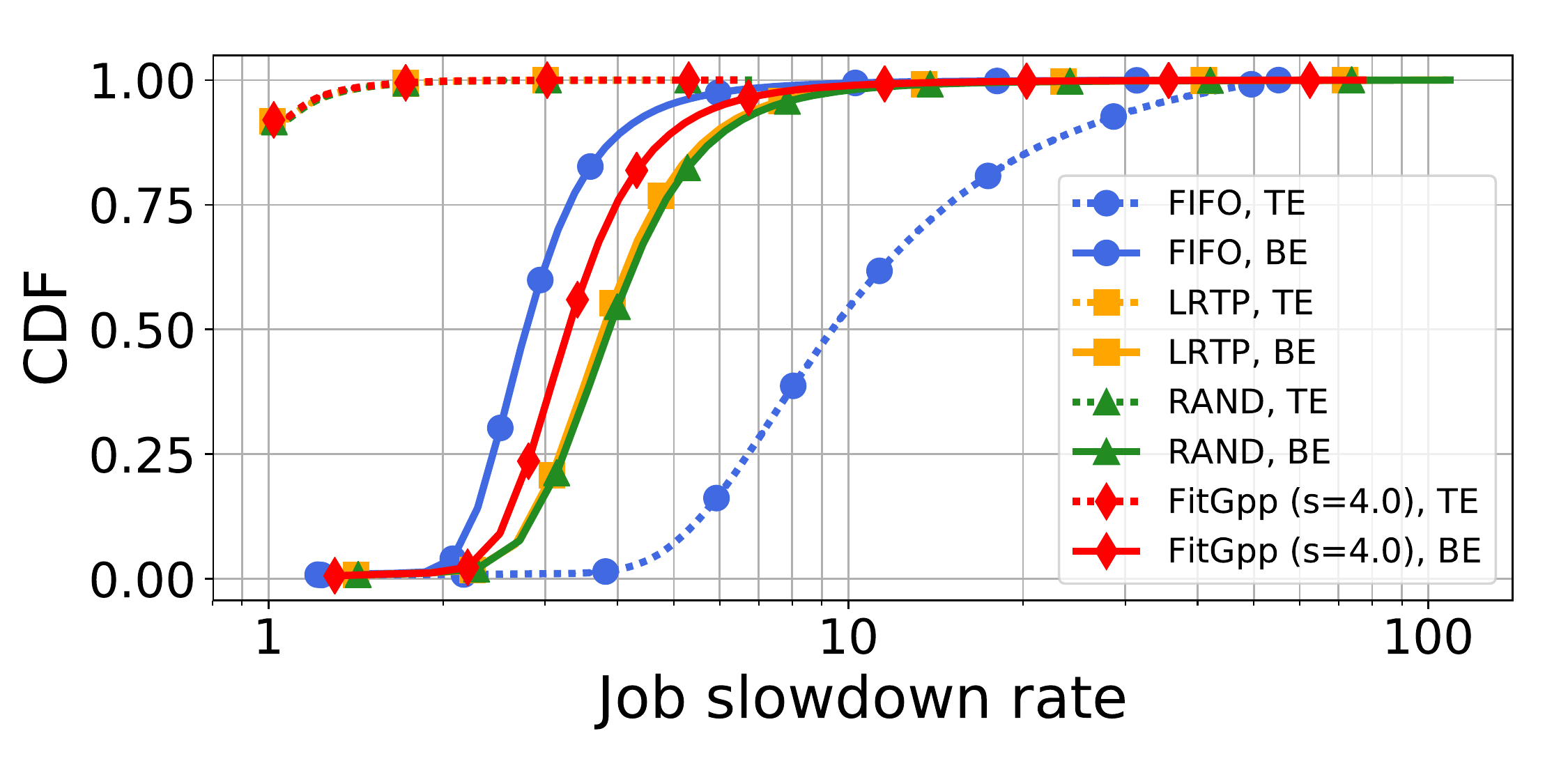}
    \caption{Job slowdown rates with synthetic workloads.}
    \label{fig:slowdown-curve}
\end{figure}

The results are given in Fig. \ref{fig:slowdown-curve} and Table \ref{tbl:slowdown-pt}.
Our FitGpp algorithm with $s = 4.0$ was able to reduce the 95th percentile of the slowdown rates of the TE jobs
by 96.6\% relative to (non-preemptive) FIFO.
Indeed, on the other hand, we expect the slowdown rate of the BE jobs in our algorithm to be worse than FIFO.
However, with our preemption strategy, FitGpp compromises the median of the slowdown rates of BE jobs by only 18.0\% and the 95th percentile by only 23.9\%.

\begin{table}[ht]
    \centering
    \caption{Percentiles of slowdown rates.}
    \begin{tabular}{lrrrrrr}
        \toprule
                                    & \multicolumn{3}{c}{TE}                        & \multicolumn{3}{c}{BE}                        \\
                                      \cmidrule(lr){2-4}                              \cmidrule(lr){5-7}
                                    & 50th          & 95th          & 99th          & 50th          & 95th          & 99th          \\
        \midrule
        FIFO                        & 9.38          & 33.4          & 48.5          & \textbf{2.78} & \textbf{4.89} & \textbf{8.21} \\
        LRTP                        & 1.00          & 1.17          & 1.58          & 3.78          & 7.25          & 12.5          \\
        RAND                        & 1.00          & 1.17          & 1.58          & 3.87          & 7.49          & 12.9          \\
        \textbf{FitGpp ($s=4.0$)}   & \textbf{1.00} & \textbf{1.15} & \textbf{1.54} & 3.28          & 6.06          & 10.3          \\
        \bottomrule
    \end{tabular}
    \label{tbl:slowdown-pt}
\end{table}

The superiority of FitGpp in this experiment is most likely due to its ability to shorten the intervals between preemptions and re-scheduling.
This is allowing our algorithm to avoid head-of-line blocking.
In fact, the median of the intervals with FitGpp is half of that of LRTP and RAND,
and the 95th percentile is 20\% shorter than LRTP and 33\% shorter than RAND (Table \ref{tbl:resched-intv}).
We shall also not forget that FitGpp makes an effort to reduce the total number of preemptions.
When $P$ (the maximum number of preemption per job) is 1,
FitGpp reduced the total number of preempted jobs to less than 7.0\% relative to LRTP and RAND (Table \ref{tbl:num-preempt-lim}).
Even when we set $P$ to be infinite,
the number of jobs preempted by FitGpp was an order of magnitude smaller than RAND and LRTP (Table \ref{tbl:num-preempt-unlim}).

\begin{table}[ht]
    \centering
    \caption{Re-scheduling intervals [min]}
    \label{tbl:resched-intv}
    \begin{tabular}{lrrrr}
        \toprule
                                    & 50th          & 75th          & 95th          & 99th          \\
        \midrule
        LRTP                        & 4.0           & 4.0           & 5.0           & 7.0           \\
        RAND                        & 4.0           & 4.0           & 6.0           & 7.0           \\
        \textbf{FitGpp ($s = 4.0$)} & \textbf{2.0}  & \textbf{2.0}  & \textbf{4.0}  & \textbf{6.0}  \\
        \bottomrule
    \end{tabular}
\end{table}

\begin{table}[H]
    \centering
    \begin{minipage}{0.39\hsize}
        \centering
        \caption{Proportion of preempted jobs (when $P = 1$).}
        \label{tbl:num-preempt-lim}
        \begin{tabular}{lr}
            \toprule
            LRTP                        & 9.6\%             \\
            RAND                        & 9.7\%             \\
            \textbf{FitGpp $(s = 4.0)$} & \textbf{6.3e-1\%} \\
            \bottomrule
        \end{tabular}
    \end{minipage}
    \hfill
    \begin{minipage}{0.59\hsize}
        \centering
        \caption{Proportion of jobs that were preempted $N$ times (when $P$ is infinite).}
        \label{tbl:num-preempt-unlim}
        \begin{tabular}{lrrr}
            \toprule
            Number of                   & \multirow{2}{*}{1}    & \multirow{2}{*}{2}    & \multirow{2}{*}{$\ge$ 3}  \\
            preemptions                 &                       &                       &                           \\
            \midrule
            LRTP                        & 6.3\%                 & 1.2\%                 & 3.4e-1\%                  \\
            RAND                        & 8.8\%                 & 4.8e-1\%              & 1.9e-2\%                  \\
            \textbf{FitGpp ($s = 4.0$)} & \textbf{5.2e-1\%}     & \textbf{3.8e-2\%}     &\textbf{9.8e-3\%}          \\
            \bottomrule
        \end{tabular}
    \end{minipage}
\end{table}

\subsection{Sensitivity analysis}
\label{subsec:sens-anal}

\begin{figure}[ht]
    \centering
    \includegraphics[width=0.85\linewidth]{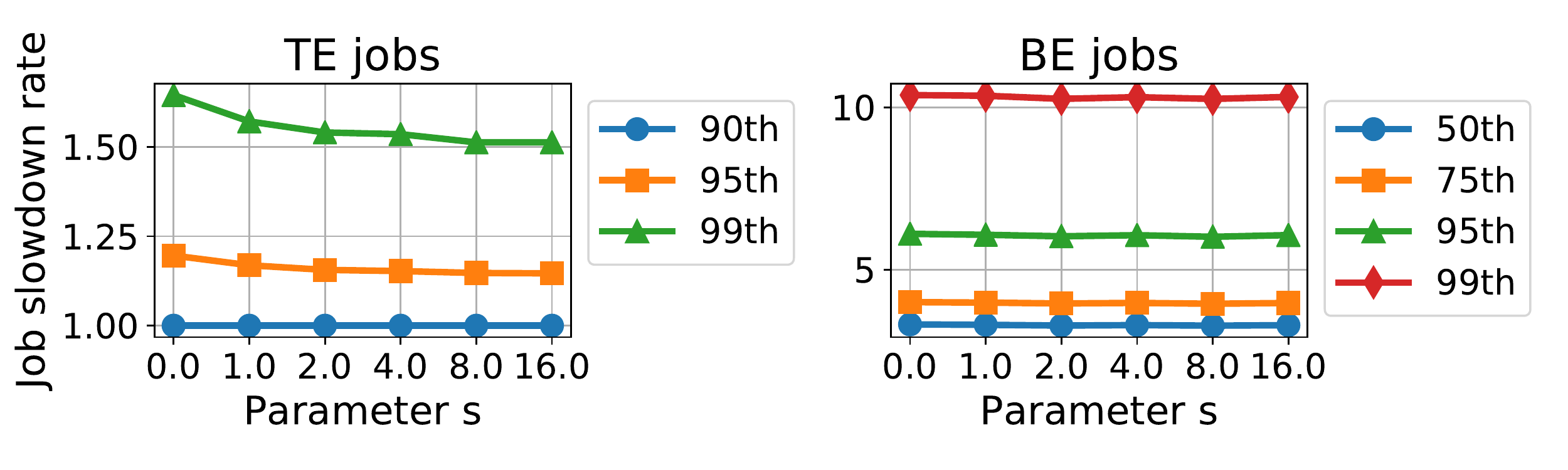}
    \caption{Job slowdown rates of FitGpp with variable value of parameter $s$.}
    \label{fig:fitgpp-param}
\end{figure}

Fig. \ref{fig:fitgpp-param} plots the percentiles of job slowdown rates as functions of FitGpp's parameter $s$ (Eq. \ref{eq:fitgpp-score}).
Recall that the value of $s$ determines importance of the length of GPs relative to the size of the resource demand.
Increasing this value makes it more likely for BE jobs with small GPs to be preempted, leading to faster scheduling of TE jobs.
The slowdown rate of TE jobs decreases as $s$ increases and saturates at some point between $s = 4.0$ and $s = 8.0$.
On the other hand, the slowdown rate of the BE jobs seems independent of the choice of $s$.

\begin{figure}[ht]
    \centering
    \includegraphics[width=0.85\linewidth]{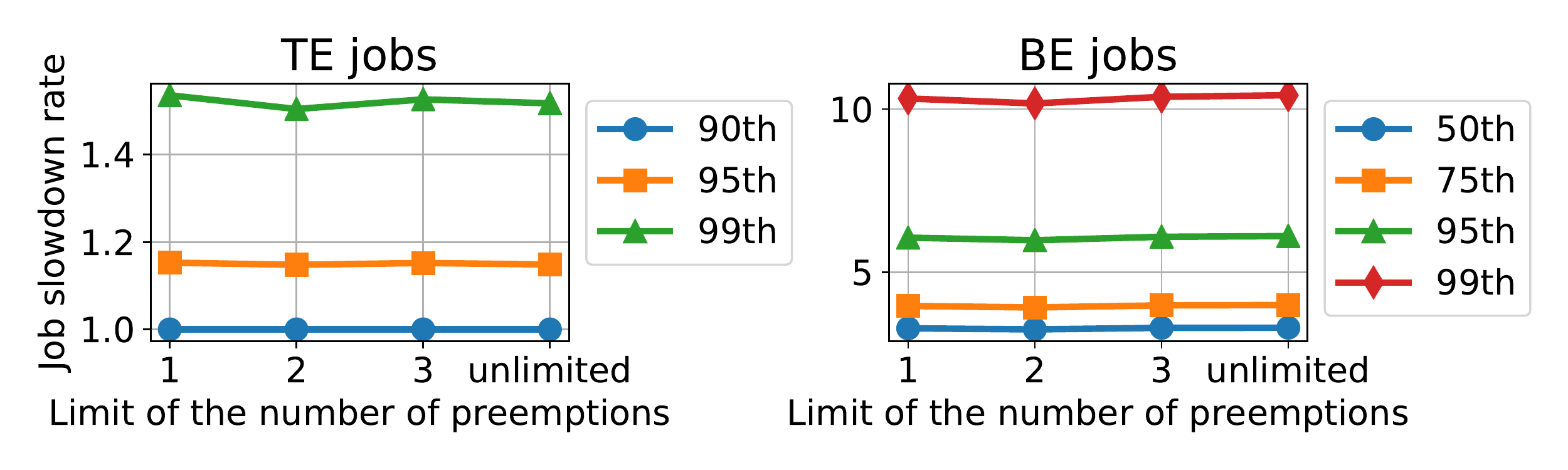}
    \caption{Job slowdown rates of FitGpp with variable value of maximum preemption limit $P$.}
    \label{fig:fitgpp-pl}
\end{figure}

Fig. \ref{fig:fitgpp-pl} plots the percentiles of job slowdown rates with various choices of $P$, the cap on the number of preemption for each job.
As we can see in the figure, the slowdown rates of both TE and BE jobs are independent of $P$.
Because FitGpp has the ability to reduce the total number of preemptions, the effect on $P$ turned out to be effectively superficial.

\begin{figure}[ht]
    \centering
    \includegraphics[width=0.85\linewidth]{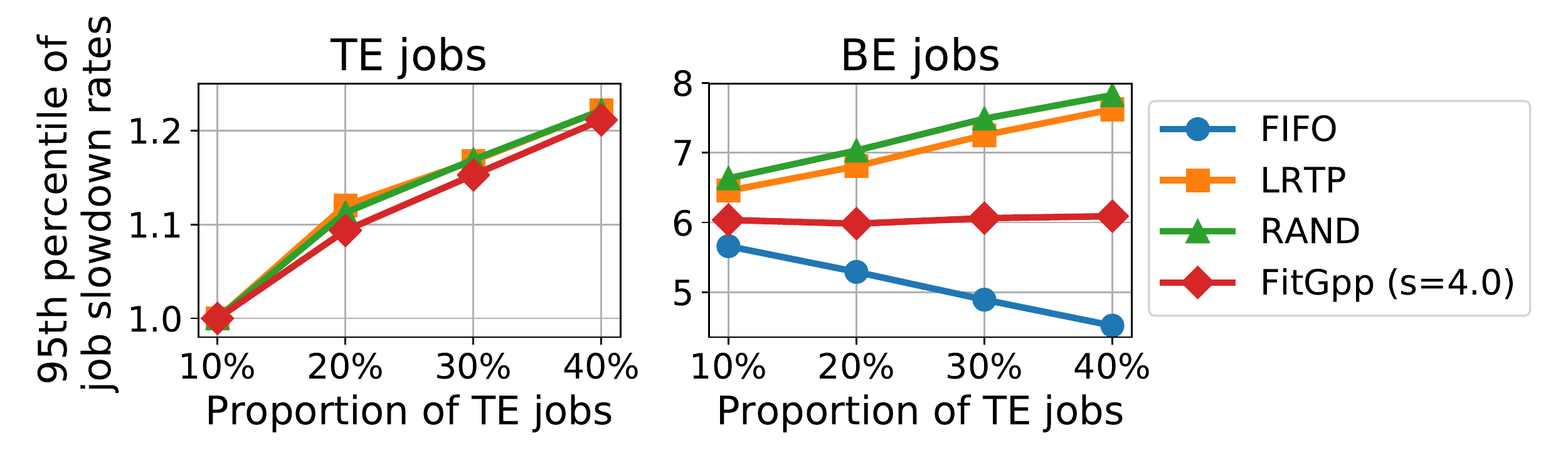}
    \caption{Job slowdown rates with variable proportion of TE jobs.}
    \label{fig:job-proportion}
\end{figure}

Fig. \ref{fig:job-proportion} plots the 95th percentile of the job slowdown rates against the proportion of the TE jobs in the workloads.
In general, the slowdown rate increases with the number of TE jobs as the sum of their resource demands begins to exceed the cluster capacity.
Note that our FitGpp with $s = 4.0$ outperforms other algorithms consistently for any proportion of the TE jobs
while maintaining the slowdown rate of the BE jobs at a relatively low level.

\begin{figure}[ht]
    \centering
    \includegraphics[width=0.85\linewidth]{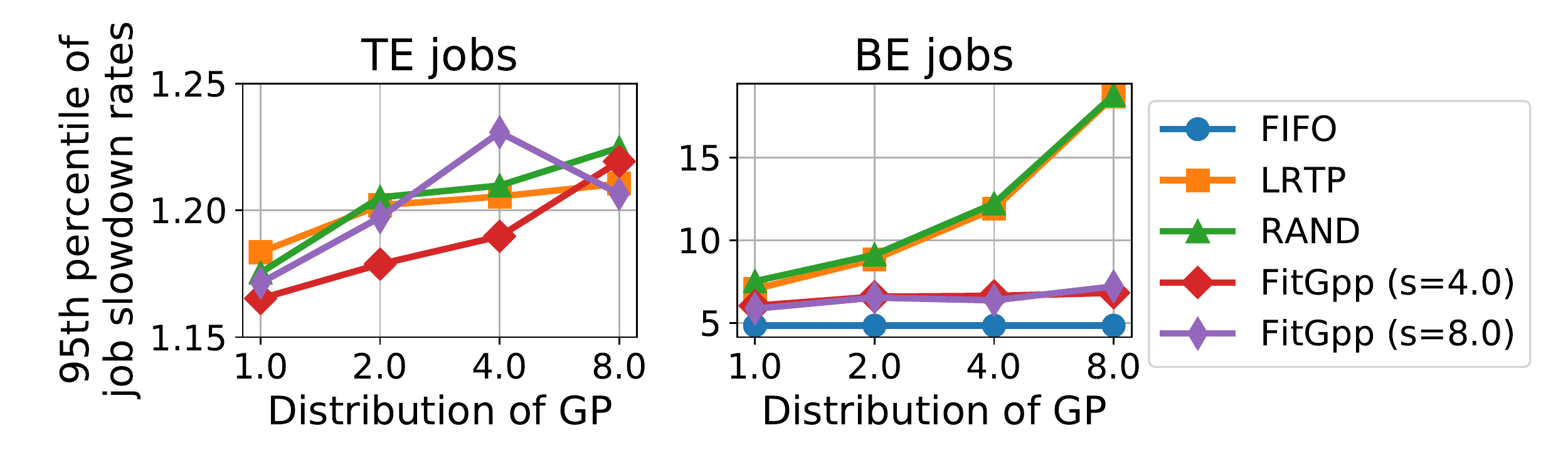}
    \caption{Job slowdown rates with variable distribution of GP length.}
    \label{fig:gp-dist}
\end{figure}

Finally, Fig. \ref{fig:gp-dist} plots the 95th percentile of the job slowdown rates against the length of GPs.
The label ``1.0''  designates the result obtained using the set of GP lengths sampled from the distribution described in Section \ref{subsec:eval-setup}.
The label ``2.0'', on the other hand, designates the result from the set of GPs sampled from the distribution
whose mean, standard deviation, and the truncation value are all twice those of the distribution used for ``1.0''.
The labels ``4.0'' and ``8.0'' designate the distributions defined analogously.
In general, the slowdown rate of the TE jobs increases with the length of GPs.
This effect can indeed be countered by choosing large $s$ value.
When the distribution of GPs is ``8.0'', FitGpp with $s = 8.0$ outperforms FitGpp with $s = 4.0$ in terms of the slowdown rate of the TE jobs.
Longer GPs also prolong the slowdown rate of the BE jobs for LRTP and RAND.
FitGpp, on the other hand, maintains the slowdown rates of the BE jobs at low values, irrespective of the value of $s$.

\subsection{Experiment with the cluster trace}
\label{subsec:cluster-trace}

\begin{figure}[ht]
    \centering
    \includegraphics[width=0.75\linewidth]{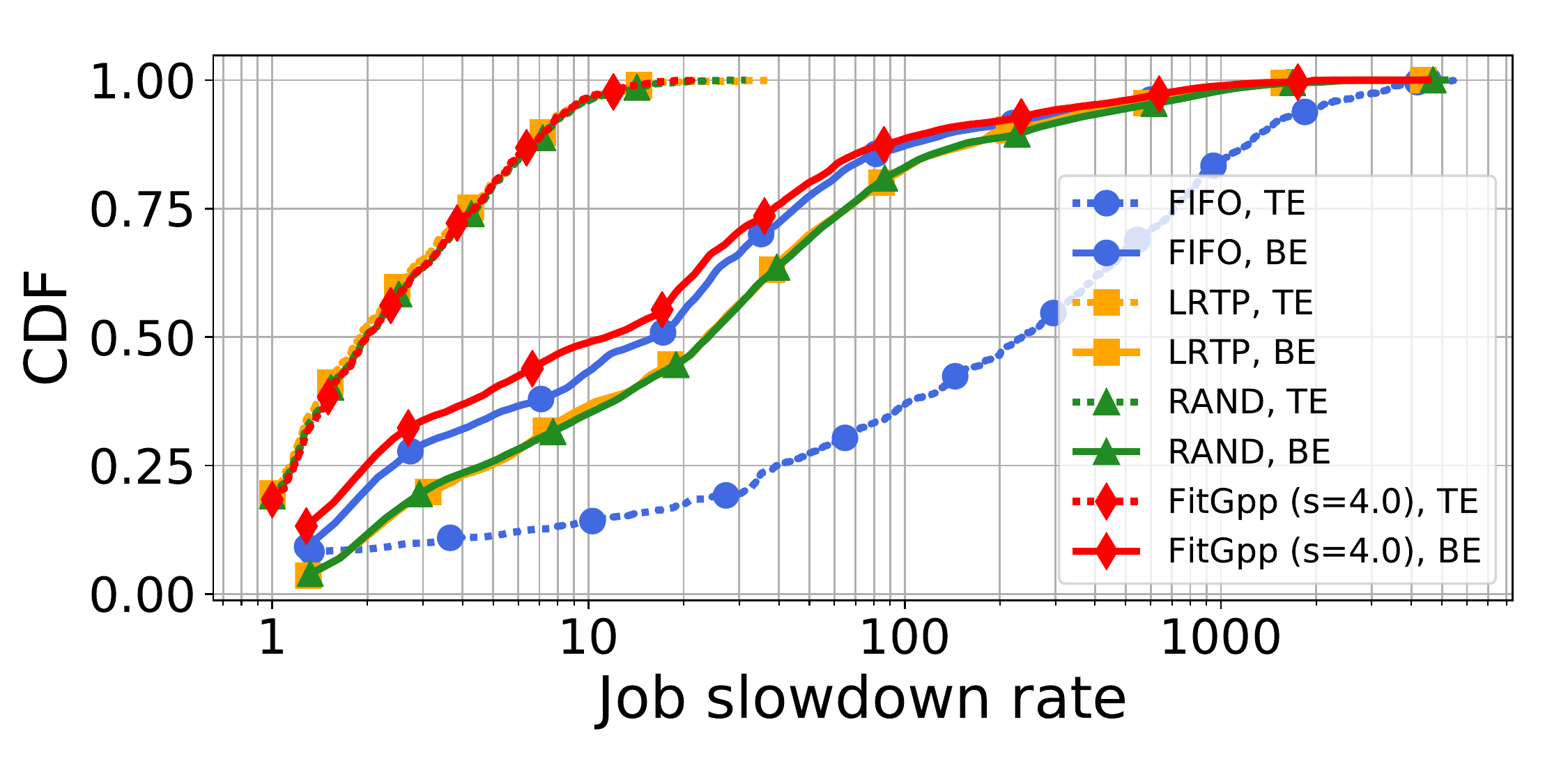}
    \caption{Job slowdown rates with the cluster trace.}
    \label{fig:dmux-slowdown-curve}
\end{figure}

We also evaluated FitGpp algorithm using the trace of the real computing cluster at the authors' institution.
Because the trace record did not contain the information regarding the length of GPs,
we had no choice but to synthesize their values by sampling from the distribution described in Section \ref{subsec:eval-setup}.
Fig. \ref{fig:dmux-slowdown-curve} and Table \ref{tbl:dmux-slowdown-pt} show the results.
Note that FitGpp with $s = 4.0$ reduces the slowdown rates of TE jobs at the 95th and 99th percentiles,
and the slowdown rates of BE jobs at the 50th, 95th, and 99th percentiles.
Depending on the situations, preemptive algorithms \textit{may} outperform non-preemptive FIFO
because the BE jobs are \textit{rearranged} in the preemption process.
In fact, the median of the slowdown rates of the BE jobs was 29.6\% lower with FitGpp relative to that of FIFO,
and the 95th percentile was 16.0\% lower with our algorithm relative to that of FIFO.

\begin{table}[H]
    \centering
    \caption{Percentiles of slowdown rates.}
    \label{tbl:dmux-slowdown-pt}
    \begin{tabular}{lrrrrrr}
        \toprule
                                    & \multicolumn{3}{c}{TE}                        & \multicolumn{3}{c}{BE}                        \\
                                      \cmidrule(lr){2-4}                              \cmidrule(lr){5-7}
                                    & 50th          & 95th          & 99th          & 50th          & 95th          & 99th          \\
        \midrule
        FIFO                        & 235           & 2080          & 3650          & 16.2          & 443           & 1180          \\
        LRTP                        & 1.90          & 9.06          & 14.5          & 23.5          & 517           & 1240          \\
        RAND                        & 1.98          & 9.18          & 15.6          & 23.9          & 559           & 1350          \\
        \textbf{FitGpp ($s=4.0$)}   & \textbf{1.98} & \textbf{9.00} & \textbf{14.2} & \textbf{11.4} & \textbf{372}  & \textbf{1050} \\
        \bottomrule
    \end{tabular}
\end{table}

\section{Conclusion}
\label{sec:conclusion}

In this paper, we presented FitGpp, a preemption algorithm that reduces the latency of the TE jobs
while reducing the slowdown of the BE jobs incurred by the preemption processes.
This study suggests the importance of systematically choosing the types of the BE job to be preempted.

Future works include the extension of this work to non-FIFO based setting, as well as the exploration of different types of score functions.
We plan to put our algorithm to test in a real environment as well.
It is also worth modifying our algorithm so that it can handle the multi-node jobs in distributed DL.
Finally, the application of our algorithm is not necessarily limited to the scheduling of DL jobs.
We shall be able to extend our algorithm to any type of workload that consists of a mixture of TE-like jobs and BE-like jobs.
It may be also interesting to apply our method to experimental designs in other scientific fields, such as biomedical science as well.

\subsubsection*{Acknowledgements}

We thank K. Uenishi, K. Fukuda, S. Maeda, and Y. Doi for fruitful discussion and reviewing this paper.
We also thank M. Koyama for a help in the composition of the paper.

\bibliographystyle{plain}
\bibliography{references}

\begin{thebibliography}{10}

\bibitem{Abadi16}
Mart{\'\i}n Abadi, Paul Barham, Jianmin Chen, Zhifeng Chen, Andy Davis, Jeffrey
  Dean, Matthieu Devin, Sanjay Ghemawat, Geoffrey Irving, Michael Isard,
  Manjunath Kudlur, Josh Levenberg, Rajat Monga, Sherry Moore, Derek~G. Murray,
  Benoit Steiner, Paul Tucker, Vijay Vasudevan, Pete Warden, Martin Wicke, Yuan
  Yu, and Xiaoqiang Zheng.
\newblock {TensorFlow: A System for Large-Scale Machine Learning}.
\newblock In {\em Proceedings of 12th {USENIX} Symposium on Operating Systems
  Design and Implementation ({OSDI} 16)}, pages 265--283, 2016.

\bibitem{Akiba17}
Takuya Akiba, Keisuke Fukuda, and Shuji Suzuki.
\newblock {ChainerMN: Scalable Distributed Deep Learning Framework}.
\newblock In {\em Workshop on ML Systems in The Thirty-first Annual Conference
  on Neural Information Processing Systems (NIPS)}, 2017.

\bibitem{Burns16}
Brendan Burns, Brian Grant, David Oppenheimer, Eric Brewer, and John Wilkes.
\newblock {Borg, Omega, and Kubernetes}.
\newblock {\em ACM Queue}, 14:70--93, 2016.

\bibitem{Chen15}
Tianqi Chen, Mu~Li, Yutian Li, Min Lin, Naiyan Wang, Minjie Wang, Tianjun Xiao,
  Bing Xu, Chiyuan Zhang, and Zheng Zhang.
\newblock {MXNet: A Flexible and Efficient Machine Learning Library for
  Heterogeneous Distributed Systems}.
\newblock In {\em NIPS Workshop on Machine Learning Systems (LearningSys)},
  2015.

\bibitem{Chen17}
Wei Chen, Jia Rao, and Xiaobo Zhou.
\newblock {Preemptive, Low Latency Datacenter Scheduling via Lightweight
  Virtualization}.
\newblock In {\em Proceedings of 2017 {USENIX} Annual Technical Conference
  ({USENIX} {ATC} 17)}, pages 251--263, 2017.

\bibitem{Delgado18}
Pamela Delgado, Diego Didona, Florin Dinu, and Willy Zwaenepoel.
\newblock {Kairos: Preemptive Data Center Scheduling Without Runtime
  Estimates}.
\newblock In {\em Proceedings of ACM Symposium of Cloud Computing conference
  (SoCC)}, pages 135--148, 2018.

\bibitem{Delgado15}
Pamela Delgado, Florin Dinu, Anne-Marie Kermarrec, and Willy Zwaenepoel.
\newblock {Hawk: Hybrid Datacenter Scheduling}.
\newblock In {\em Proceedings of 2015 {USENIX} Annual Technical Conference
  ({USENIX} {ATC} 15)}, pages 499--510, 2015.

\bibitem{Paszke17}
Adam Paszke, Sam Gross, Soumith Chintala, Gregory Chanan, Edward Yang, Zachary
  DeVito, Zeming Lin, Alban Desmaison, Luca Antiga, and Adam Lerer.
\newblock {Automatic differentiation in PyTorch}.
\newblock In {\em Autodiff Workshop in The Thirty-first Annual Conference on
  Neural Information Processing Systems (NIPS)}, 2017.

\bibitem{Peng18}
Yanghua Peng, Yixin Bao, Yangrui Chen, Chuan Wu, and Chuanxiong Guo.
\newblock {Optimus: An Efficient Dynamic Resource Scheduler for Deep Learning
  Clusters}.
\newblock In {\em Proceedings of Thirteenth EuroSys Conference ({EuroSys}
  '18)}, 2018.

\bibitem{Vavilapalli13}
Vinod~Kumar Vavilapalli, Siddharth Seth, Bikas Saha, Carlo Curino, Owen
  O'Malley, Sanjay Radia, Benjamin Reed, Eric Baldeschwieler, Arun~C Murthy,
  Chris Douglas, Sharad Agarwal, Mahadev Konar, Robert Evans, Thomas Graves,
  Jason Lowe, and Hitesh Shah.
\newblock {Apache Hadoop YARN: Yet Another Resource Negotiator}.
\newblock In {\em Proceedings of the 4th annual Symposium on Cloud Computing
  (SoCC)}, 2013.

\bibitem{Xiao2018}
Wencong Xiao, Romil Bhardwaj, Ramachandran Ramjee, Muthian Sivathanu, Nipun
  Kwatra, Zhenhua Han, Pratyush Patel, Xuan Peng, Hanyu Zhao, Quanlu Zhang, Fan
  Yang, and Lidong Zhou.
\newblock {Gandiva: Introspective Cluster Scheduling for Deep Learning}.
\newblock In {\em 13th {USENIX} Symposium on Operating Systems Design and
  Implementation ({OSDI} 18)}, pages 515--610, 2018.

\bibitem{Zaharia10}
Matei Zaharia, Mosharaf Chowdhury, Michael~J. Franklin, Scott Shenker, and Ion
  Stoica.
\newblock {Spark: Cluster Computing with Working Sets}.
\newblock In {\em Proceedings of HotCloud 2010}, 2010.

\end{thebibliography}

\end{document}